\documentclass[twocolumn]{webofc}
\usepackage[varg]{txfonts}   
\usepackage{bm}
\usepackage{amsmath}
\usepackage{physics}
\usepackage{comment}
\usepackage{lipsum}
\usepackage{float}
\usepackage{fancyhdr}
\begin{document}
\title{Statistical sensitivity of the nEDM apparatus at PSI to $n-n'$ oscillations}
%
%

\author{
\firstname{C.} \lastname{Abel}\inst{1}\and 
\firstname{N.~J.} \lastname{Ayres}\inst{1}\and 
\firstname{G.} \lastname{Bison}\inst{2}\and 
\firstname{K.} \lastname{Bodek}\inst{3}\and 
\firstname{V.} \lastname{Bondar}\inst{4}\and 
\firstname{P.-J.} \lastname{Chiu}\inst{2,5}\and 
\firstname{M.} \lastname{Daum}\inst{2}\and 
\firstname{S.} \lastname{Emmenegger}\inst{5}\and 
\firstname{P.} \lastname{Flaux}\inst{6}\and 
\firstname{L.} \lastname{Ferraris-Bouchez}\inst{7}\and 
\firstname{W.~C.} \lastname{Griffith}\inst{1}\and 
\firstname{Z.~D.} \lastname{Gruji\'c}\inst{8}\and 
\firstname{N.} \lastname{Hild}\inst{2,5}\and 
\firstname{K.} \lastname{Kirch}\inst{2,5}\and 
\firstname{P.~A.} \lastname{Koss}\inst{4}\and 
\firstname{A.} \lastname{Kozela}\inst{9}\and 
\firstname{J.} \lastname{Krempel}\inst{5}\and 
\firstname{B.} \lastname{Lauss}\inst{2}\and 
\firstname{T.} \lastname{Lefort}\inst{6}\and 
\firstname{A.} \lastname{Leredde}\inst{7}\and 
\firstname{P.} \lastname{Mohanmurthy}\inst{2,5}\thanks{\email{prajwal@mohanmurthy.com}}\and 
\firstname{O.} \lastname{Naviliat-Cuncic}\inst{6}\thanks{now at Michigan State University, East-Lansing, MI 48824, USA}\and 
\firstname{D.} \lastname{Pais}\inst{2,5}\and 
\firstname{F.~M.} \lastname{Piegsa}\inst{10}\and 
\firstname{G.} \lastname{Pignol}\inst{7}\and 
\firstname{M.} \lastname{Rawlik}\inst{5}\and 
\firstname{D.} \lastname{Rebreyend}\inst{7}\and 
\firstname{D.} \lastname{Ries}\inst{11}\and 
\firstname{S.} \lastname{Roccia}\inst{12} \thanks{present address: Institut Laue Langevin, 38000 Grenoble, France}\and 
\firstname{D.} \lastname{Rozpedzik}\inst{3}\and 
\firstname{P.} \lastname{Schmidt-Wellenburg}\inst{2}\and 
\firstname{A.} \lastname{Schnabel}\inst{13}\and 
\firstname{N.} \lastname{Severijns}\inst{4}\and 
\firstname{J.} \lastname{Thorne}\inst{10}\and 
\firstname{R.} \lastname{Virot}\inst{7}\and 
\firstname{J.} \lastname{Zejma}\inst{3}\and 
\firstname{G.} \lastname{Zsigmond}\inst{2}\thanks{\email{geza.zsigmond@psi.ch}}
}

\institute{
University of Sussex, Brighton BN1 9RH, United Kingdom
\and
Paul Scherrer Institute, 5232 Villigen, Switzerland
\and
Marian Smoluchowski Institute of Physics, Jagiellonian University, 30-348 Krak\'ow, Poland
\and
Institute for Nuclear and Radiation Physics, KU Leuven, 3001 Heverlee, Belgium 
\and
Institute for Particle Physics and Astrophysics, ETH Z{\"u}rich, 8093 Z{\"u}rich, Switzerland 
\and
LPC Caen, ENSICAEN, Normandie Universit\'e, CNRS/IN2P3, 14000 Caen, France
\and
Laboratoire de Physique Subatomique et de Cosmologie, 38000 Grenoble, France
\and
University of Fribourg, 1700 Fribourg, Switzerland 
\and
Henryk Niedwodnicza\'nski Institute of Nuclear Physics, 31-342 Krak\'ow, Poland
\and
Laboratory for High Energy Physics and Albert Einstein Center for Fundamental Physics, University of Bern, 3012 Bern, Switzerland 
\and
Institut f{\"u}r Kernchemie, Johannes Gutenberg-Universit{\"a}t, 55128 Mainz, Germany
\and
CSNSM, Universit\'e Paris Sud, CNRS/IN2P3, Universit\'e Paris Saclay, 91405 Orsay-Campus, France
\and
Physikalisch Technische Bundesanstalt, 10587 Berlin, Germany
}

\abstract{%
The neutron and its hypothetical mirror counterpart, a sterile state degenerate in mass, could spontaneously mix in a process much faster than the neutron $\beta$-decay. Two groups have performed a series of experiments in search of neutron - mirror-neutron ($n-n'$) oscillations. They reported no evidence, thereby setting stringent limits on the oscillation time $\tau_{nn'}$. Later, these data sets have been further analyzed by Berezhiani et al.(2009-2017), and  signals, compatible with $n-n'$ oscillations in the presence of mirror magnetic fields, have been reported. The Neutron Electric Dipole Moment Collaboration based at the Paul Scherrer Institute performed a new series of experiments to further test these signals. In this paper, we describe and motivate our choice of run configurations with an optimal filling time of $29~$s, storage times of $180~$s and $380~$s, and applied magnetic fields of $10~\mu$T and $20~\mu$T. The choice of these run configurations ensures a reliable overlap in settings with the previous efforts and also improves the sensitivity to test the signals. We also elaborate on the technique of normalizing the neutron counts, making such a counting experiment at the ultra-cold neutron source at the Paul Scherrer Institute possible. Furthermore, the magnetic field characterization to meet the requirements of this $n-n'$ oscillation search is demonstrated. Finally, we show that this effort has a statistical sensitivity to $n-n'$ oscillations comparable to the current leading constraints for $B'=0$.
}
\maketitle
\section{Introduction}
\label{intro}
In the paper in which Lee and Yang proposed parity violation \cite{[6-1]}, they also noted that it may be resolved by the introduction of a parity conjugated copy of standard model (SM) particles - SM$^{\prime}$. Kobzarev, Okun and Pomeranchuk further developed this idea and formalized mirror matter \cite{[1]}. They also introduced mirror photons, \emph{i.e.} mirror magnetic fields which were not defined in Ref. \cite{[6-1]}. Foot and Volkas showed that with the introduction of mirror matter, parity and time reversal symmetries could be restored in the global sense \cite{[6-1-1],[6-1-2]}. 

Mixing of SM$^{\prime}$ and SM particles could provide answers for several long-standing issues in physics today. Mirror matter could act as a dark matter candidate \cite{[6-2-0],[6-2-1],[6-2-2],[6-2-3],[6-2-4],[6-2-5],[6-2-6]}. It could provide a mechanism to help solve the sterile neutrino anomaly \cite{[6-3-1],[6-3-2],[6-3-3]}. Mirror matter could also provide an additional channel of CP violation through mixing of SM and SM$^{\prime}$ particles, thus helping to explain baryogenesis \cite{[6-4]}. It could provide a mechanism to relax the Greisen-Zatsepin-Kuzmin (GZK) limit through $n-n'$ oscillations \cite{[6-5-1],[6-5-2]}. Baryogenesis also requires baryon number violation. Neutron - mirror-neutron oscillations are one such process \cite{[6-5-3]}. The corresponding oscillation time may be smaller than the neutron $\beta-$decay lifetime \cite{[6-5-4]}. A historical overview of the physics of mirror matter can be found in Ref. \cite{[6-6]}.

It was shown in Ref.~\cite{[4]} that neutrons may mix with mirror-neutrons with an interaction Hamiltonian:
\begin{eqnarray}
\mathcal{H} = \begin{bmatrix}
\mu_n \bm{B \cdot \sigma} & \hbar/\tau_{nn'} \\
\hbar/\tau_{nn'} & \mu_n \bm{B' \cdot \sigma}
\end{bmatrix}\label{eq5-6},
\end{eqnarray}
where $\tau_{nn'}$ is the $n-n'$ oscillation time, and ${\bf B}$,~${\bf B^{'}}$ are the magnetic and mirror magnetic fields, respectively. Eq.~\ref{eq5-6} shows that applying a magnetic field (${\bf B'} \ne {\bf B}$) lifts the degeneracy between neutron and mirror-neutron states, thereby suppressing the oscillation between the states. From Eq. \ref{eq5-6} it follows that there can be two general techniques for searching for such oscillations:
\begin{enumerate}
\item{ultra-cold neutron (UCN) storage experiments, where one searches for a magnetic field dependence of the storage curve \cite{[2],[3],[5]},}
\item{regeneration experiments, where a “particle through a wall” measurement is performed to look for cold neutrons regenerating after crossing a barrier in `mirror state' \cite{[5-1],[5-2]}.}
\end{enumerate}

Our attempt to search for $n-n'$ oscillations employs the UCN storage technique. Counting the number of stored neutrons as a function of time ($N(t_s)$) is usually referred to as the storage curve. We performed an experiment where the storage curves of neutrons were compared with magnetic field turned on and off. Neutron - mirror-neutron oscillation, depending on the magnetic field(s) applied, may add an additional loss channel to the storage curve:
\begin{eqnarray}
N_{\{0, B\}}(t_s) &=& N_{\{0, B\}}(t_s=0) \\\nonumber &&\times\exp{-\left(R + R^{nn'}_{\{0, B\}}\right)t_s} \label{eq5-32m}.
\end{eqnarray}
Here, $N_{\{0, B\}}(t_s=0)$ are the initial numbers of neutrons stored in the chamber and  $N_{\{0, B\}}(t_s)$ denotes the UCN data points as a function of storage time $t_s$, when the magnetic field is switched off or switched on, respectively. The value $R$ refers to the sum of rates of known loss channels as up-scattering, absorption, and $\beta-$decay of neutrons in the chamber, which are all independent of the applied magnetic field.  $R^{nn'}_{\{0, B\}}$ is the loss channel added due to $n-n'$ oscillations when the magnetic field applied is zero or when it is different from zero. Note that when the applied magnetic field, $B$, is zero, there could still be a non-zero mirror magnetic field, $B'\ne0$. The additional loss channel due to $n-n'$ oscillation can be isolated as derived in section 3 of \cite{[4]}.  By taking the ratio of UCN counts with magnetic field turned off ($B_0\sim0$) and on ($B_0>0$) and the decay rate formulas in \cite{[4]} follows:
\begin{eqnarray}
E_B\left(t_s\right) &=& \frac{N_0\left(t_s\right)}{N_B\left(t_s\right)} - 1\label{eq5-32b}\\
&=& \exp{-\left(R^{nn'}_0 - R^{nn'}_B\right)t_s} - 1\label{eq5-32b2}\\
&=& \frac{t_s}{\left<t_f\right>}\frac{\eta^2\left(3-\eta^2\right)}{2\tau_{nn'}^2\omega'^2\left(1-\eta^2\right)^2} \label{eq5-32},
\end{eqnarray}
where $\omega^{(')} = 45.81~(\mu\text{T.s})^{-1}\cdot B^{(')}$, $\eta = B/B'$, and $\left<t_f\right>$ is the mean time between two consecutive wall collisions of the UCN within the storage chamber. Eq. \ref{eq5-32} is valid in the range $\omega' \left<t_f\right> \gg 1$  \cite{[4]}. In the case when no mirror magnetic field is assumed, Eq. \ref{eq5-32b2} reduces to:
\begin{eqnarray}
E_0(t_s)  &\approx& \frac{\left<t^2_f\right>}{\left<t_f\right>}\frac{t_s}{\tau^2_{nn'}}.\label{eq5-13}
\end{eqnarray}
Eq. \ref{eq5-13} is valid in the range $\omega \left<t_f\right> \ll 1$. Eqs. \ref{eq5-32} and \ref{eq5-13} form the ``ratio channel'' of $n-n'$ oscillations. To summarize, $E_B$ denotes the generalized case when $B'\ne0$, whereas $E_0$ denotes the special case when we assume $B'=0$. 

In Ref.~\cite{[4]}, Berezhiani showed that the asymmetry between storage curves is also sensitive to $n-n'$ oscillations when magnetic fields of opposing directions are applied:
\begin{eqnarray}
A_B\left(t_s\right) &=& \frac{\left(N_{\bf B}\left(t_s\right)-N_{\bf{-B}}\left(t_s\right)\right)}{\left(N_{\bf B}\left(t_s\right)+N_{\bf{-B}}\left(t_s\right)\right)} \nonumber \\
&=& -\frac{t_s}{\left<t_f\right>}\frac{\eta^3\cos\left(\beta\right)}{\tau^2_{nn'}\omega^2\left(1-\eta^2\right)^2}. \label{eq5-34}
\end{eqnarray}
Here the new variable $\beta$ is the angle between ${\bf B}$ and ${\bf B'}$. Eq. \ref{eq5-34} forms the ``asymmetry channel'' of $n-n'$ oscillations. Particularly, the asymmetry channel is sensitive to the direction of the mirror magnetic field \emph{w.r.t} the applied magnetic field. 

If there are no $n-n'$ oscillations, the measured values of $E_B$, $E_0$, $A_B$ would all be zero. This defines the null hypothesis. In order to measure $n-n'$ oscillations, searches for deviations from the null-hypothesis using both the UCN storage technique, as well as the regeneration technique, have been attempted.

The first experiments in search of $n-n'$ oscillations were performed using the ratio channel. They set the limits $\tau_{nn'} >103~\mbox{s (95 \% C.L.)}$ \cite{[2]} and $\tau_{nn'} > 414~\mbox{s (90 \% C.L.)}$ \cite{[3]}, respectively. The current leading limit on the oscillation time under the assumption the mirror magnetic field, $B'=0$ is $\tau_{nn'} > 448~\mbox{s (90 \% C.L.)}$ \cite{[3-2]}. The current leading constraint for the case of $B'\ne0$ is $\tau_{nn'} > 12~\mbox{s (95\% C.L.) over}~(0 < B' < 12.5~\mu$T$)$ \cite{[5]}. Reanalysis of experiments in Refs. \cite{[2],[3],[5]} by Berezhiani et al. in  \cite{[4],[4-2],[6]} showed signals in the asymmetry channel in the region $6 < B' < 40~\mu$T, which could not be excluded by any experiment thus far. Testing these signals was the primary motivation for the new side-project of the Neutron Electric Dipole Moment (nEDM) Collaboration \cite{[6-0]} searching for neutron - mirror-neutron oscillations with the apparatus at the Paul Scherrer Institute (PSI).

Our search made use of the PSI UCN source \cite{[7-0-1-1],[7-0-1-2],[7-0-1-3],[7-0-1-4],[8-0]} and a re-purposed nEDM apparatus \cite{[10-2],[7-1-1-1]} for this experiment. At PSI neutrons are produced from a proton driven spallation source. These spallation neutrons are then moderated in heavy water and successively down-scattered in solid deuterium to UCNs. In the PSI nEDM apparatus, a switch directs the neutrons to and from the storage chamber. The neutrons can be stored under the influence of magnetic (and electric) fields. The storage chamber is enclosed in a 4-layer $\mu$-metal shield which is housed inside an active magnetic-field compensation system \cite{[7-0-2]}. A cycle usually involves letting the UCNs from the source fill the storage chamber after passing through the appropriately configured switch. The UCN shutter at the bottom of the storage chamber then closes and the storage of UCNs begins. During this time, the rest of the UCN exiting from the source are guided to the detectors and are used as monitor counts. After a period of storage, the UCN shutter below the storage chamber is opened and the neutrons are spin-analyzed in the U-shaped Simultaneous Spin Analyzer (USSA) \cite{[7-1-1]} before being simultaneously detected by a pair of neutron detectors (called NANOSC-A and B), each detecting a single spin state \cite{[7-1-2]}. A schematic diagram of the PSI nEDM apparatus can be found in Ref. \cite{[7-1-1-1]}. The cycle schedule is handled by a slow-control software called `micro-timer'. In the mirror-neutron search, we used unpolarized neutrons, thus spin-analysis was not required. Patterns of different magnetic fields were applied after every 4 cycles, referred to in the following as a `dwell'.

\section{Determining the effective storage time}
\begin{figure}[h!]
\centering
\includegraphics[scale=.31]{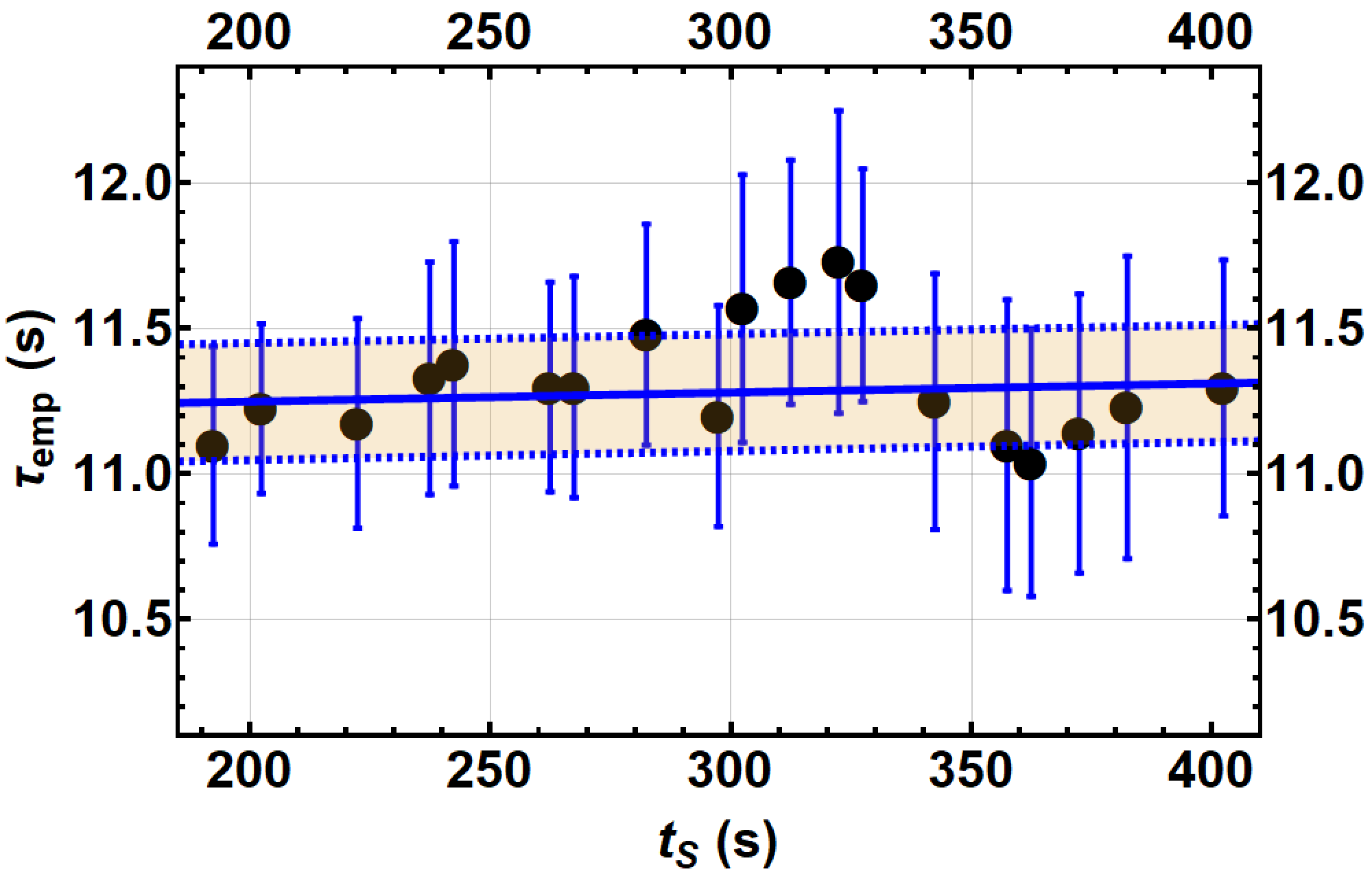}
\caption[]{The emptying time constant of UCNs as a function of micro-timer storage time. The solid line indicates a linear fit. The dashed lines show the case for a $\pm 1\sigma$ deviation of the constant term.}
\label{fig5-16}
\end{figure}

The emptying phase is the period of time when the UCNs in the nEDM storage chamber are allowed to drop down into the detector systems and be detected after a period of storage. During the emptying phase, the UCN shutter at the bottom of the nEDM storage chamber is opened and the switch is set to the empty position. During this phase, time-of-arrival spectra are recorded in the UCN detectors. 
A set of emptying time-constants $\tau_{\text{emp}}$ can be extracted by fitting these spectra (also called `emptying curves') in the time-of-arrival histograms with an exponential-decay function. These fits yield cycle-by-cycle values for $\tau_{\text{emp}}$ as a function of micro-timer storage time. These values averaged over all cycles for a given storage time are shown in Fig.~\ref{fig5-16}. The corresponding errors on the mean do not reflect the large scatter of the central values shown in Fig.~\ref{fig5-16}. This is caused by the fluctuations in the motion of the switch, and in the opening and closing of the UCN shutter at the bottom of the UCN chamber. Therefore in Fig.~\ref{fig5-16} the standard deviations are shown as error bars. 

Studying the emptying time-constant also allows us to understand the effective storage time of neutrons which is different from the micro-timer storage time for the following reasons. 
The neutrons experience the same magnetic field, which is present during storage, also during both the filling and the emptying phases when neutrons could oscillate into their mirror counterparts as well. While the detected neutrons are recorded with a time tag (with a precision of $1$~ns), it is hard to precisely say when the neutrons left the nEDM storage chamber once the UCN shutter at its bottom was opened. Also, it is not possible to tell the precise time at which a neutron entered the nEDM storage chamber during the filling phase. This makes the micro-timer storage time $t^*_s$ not equal to the precise time period during which they were present in the chamber. The micro-timer storage time starts with the end of the filling phase and ends at the beginning of the emptying phase, and it is indicated by $t^*_s$ here. 

The analysis using Eqs.~\ref{eq5-32}, \ref{eq5-13} and \ref{eq5-34} relies on the mean time (the average over UCN trajectories) the UCN have spent in the magnetic field. As explained above, this should also include the  mean time of emptying and filling. The mean emptying time is given by the time-constant $\tau_{\text{emp}}$ since the time-spectra read by the detectors have a shape of an exponential decay. 
The chamber is also filled through the same opening, and same vertical guide (narrower than the beam-line guides), as it is emptied. Thus, for a given energy spectrum, the mean filling time  will be approximately equal to the mean emptying time. The optimal filling time, a different quantity discussed in the next section, is about three times larger. 
Only the UCNs that survived the storage phase and were counted are relevant. These surviving UCNs would thus have similar filling and emptying time constants, owing to the fact that the surviving UCNs would have had similar energy spectra during both the filling and emptying phases. Consequently, the effective time of storage can be obtained by adding an additional time period, $2\tau_{\text{emp}}$, representing contributions of the mean times of filling and emptying, to the micro-timer storage time as follows:
\begin{eqnarray}
t_s &=& t^*_s + 2\tau_{\text{emp}} . \label{eq5-41b}
\end{eqnarray}

The emptying time is a function of UCN velocity. Furthermore, we cannot exclude that the energy spectra of the neutrons after storage could change when varying the storage time because faster UCN are lost at a higher rate than slower ones. 
Taking the data shown in Fig.~\ref{fig5-16}, we can assume that if there is a dependence of $\tau_{\text{emp}}$ on the storage time, this is linear to a first approximation.  
Henceforth, the effective `storage-time' will imply adding to the micro-timer storage-time $2\tau_{\text{emp}}$ from the linear fit, and the associated fit error.

\section{Optimizing the filling time}
\begin{figure}[h!]
\centering
\includegraphics[scale=.45]{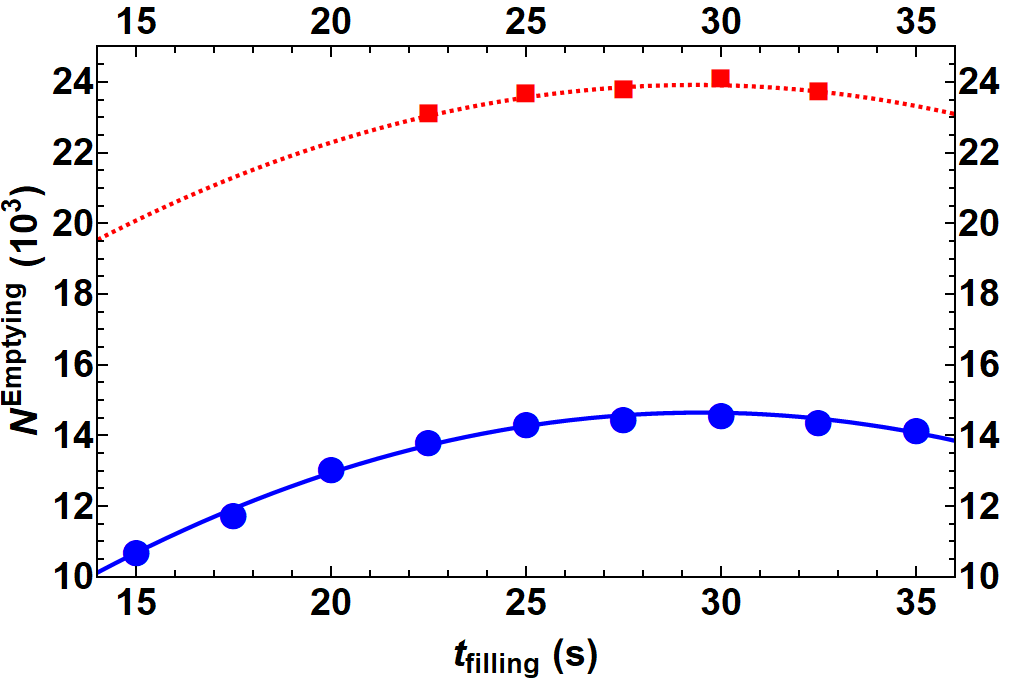}
\caption[]{Plot showing the variation of total neutron counts after a fixed period of storage \emph{w.r.t.} the filling time in seconds, along with a parabolic fit to the data-set. The red data set, with squares for data points and dashed line fit, corresponds to $t^*_s = 90~$s, whereas the blue data set, with dots for data points and solid line fit, corresponds with $t^*_s = 180~$s.}
\label{fig5-13}
\end{figure}
The filling time is defined as the time interval after a proton beam pulse during which the UCN shutter at the bottom of the nEDM storage chamber is open, while the UCN switch is set to the filling position. As UCN were only produced during a short period of $7.5~$s in the source, the number of neutrons delivered to the storage cell decreases once production stops. At the same time UCNs also leave the nEDM cell during the same filling period, or some of them are lost on the walls, similarly as  during storage. The optimal filling time is the moment when filling and loss rates are equal. As mentioned earlier, the storage time determines the energy spectrum of those detected UCN which have been filled. 

We studied the optimal filling time by varying it while keeping the storage time constant. Figure~\ref{fig5-13} shows two curves corresponding to the case where the micro-timer storage time was held fixed at $90$ s and $180$ s respectively, while the filling time was scanned. A parabolic fit to the data yields an optimal filling time: 
\begin{eqnarray}
t^{\text{optimal}}_{\text{filling}}(t^*_s = 180~\text{s}) &=& 29.4 \pm 0.1~\text{s}, \label{eq5-39} \\
t^{\text{optimal}}_{\text{filling}}(t^*_s = 90~\text{s}) &=& 29.3 \pm 0.1~\text{s}. \label{eq5-40}
\end{eqnarray}
We notice that within the error bars, the optimal filling time is independent of the UCN storage time. We used $29~$s as the filling time for all the runs in this effort.

\begin{figure*}[h!]
\centering
\includegraphics[scale=.1475]{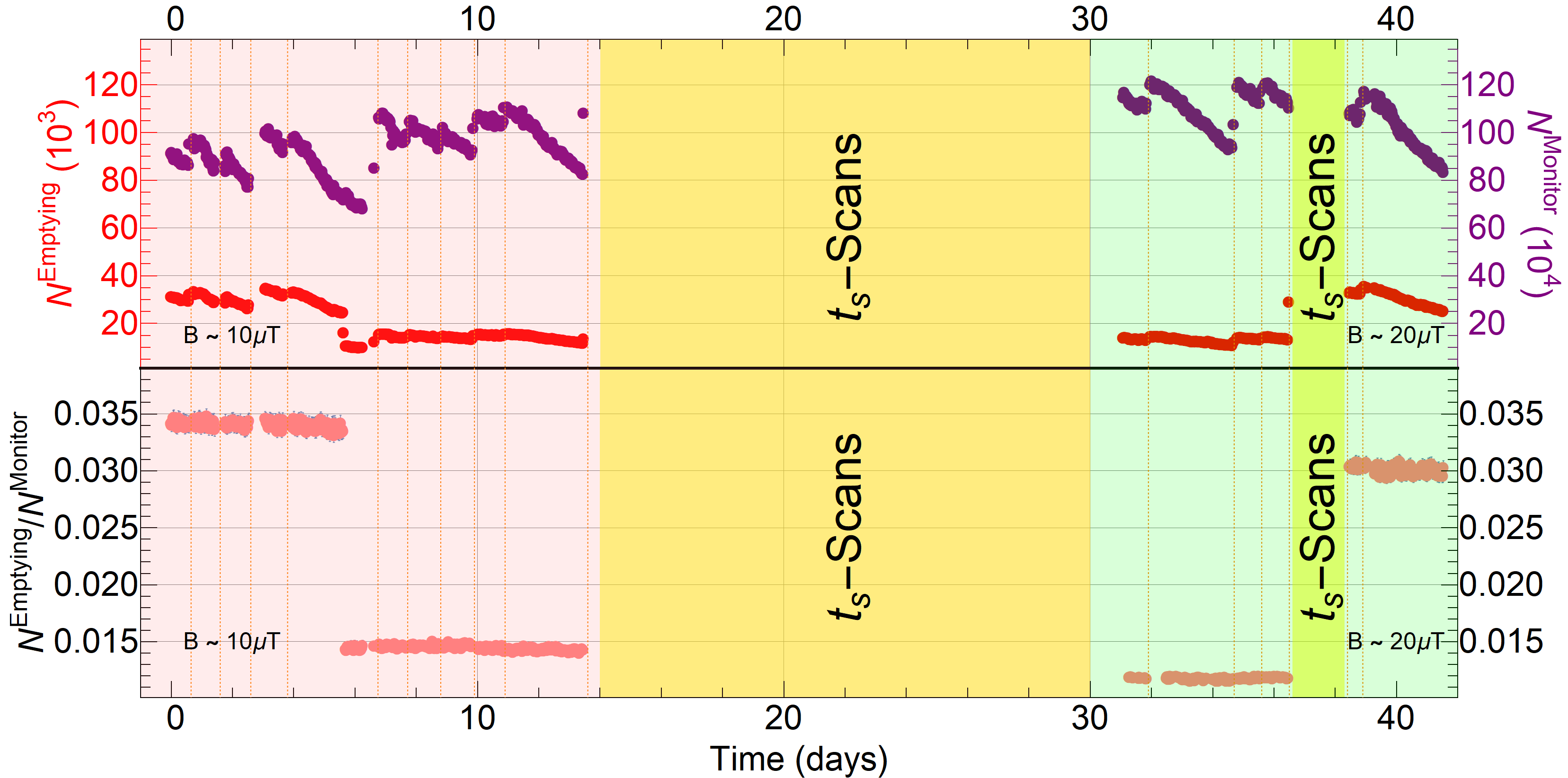}
\caption{(Top) Plot showing UCN monitor counts in purple (in units of ten thousands, corresponding to the vertical axis on the right hand side), and emptying counts in red (in units of thousands, corresponding to the vertical axis on the left hand side), both averaged over a dwell of 4 cycles over the entire data taking stretch of the $n-n'$ oscillation search. (Bottom) Ratio between the emptying and monitor counts as a function of time, for each dwell of 4 cycles. In both plots, yellow regions labeled - `$t_s-$Scans' refer to a time period when the stored UCN energy spectra were estimated. The pink regions labeled `$B\sim10~\mu$T'  and green regions labeled `$B\sim20~\mu$T' refer to runs which were collected by applying the corresponding indicated magnetic field. The large steps in the normalized counts correspond to switching between the micro-timer storage times, 180~s and 380~s. The vertical orange dotted lines indicate when the solid deuterium moderator was reconditioned \cite{[8-0]}, to recover the maximal neutron flux from the UCN source.}
\label{fig5-17}
\end{figure*}
\section{Normalizing emptying counts with monitor counts}
So far, a common feature of sD$_2$ based UCN sources at neutron spallation targets is that the maximum intensity, after each proton pulse, decays over timescales of the order of days \cite{[8-0]}, as shown in Fig.~\ref{fig5-17} (top). In order to use Eqs. \ref{eq5-32} and \ref{eq5-34}, one has to assume that the number of neutrons that were filled until the beginning of the storage time is stable. If that is not the case, the counts have to be normalized by a reliable method. This allows us to use counts such as $N_0(t_s)$, $N_{\bf B}(t_s)$, and $N_{-{\bf B}}(t_s)$ from different cycles. Other experiments searching for mirror-neutrons based on UCN counting, such as Ban~et~al.~(2007) \cite{[2]}, Serebrov~et~al.~(2008) \cite{[3]}, and Altarev~et~al.~(2009) \cite{[5]}, used neutrons from the PF2 source at the Institute Laue-Langevin (ILL) \cite{Steyerl86}, where the initial neutron counts did not decay over time. The initial number of neutrons in the ILL turbine-based experiments only fluctuated by about $2$\% over time correlating to the reactor performance. Here, we present our method to correct for both fluctuations and drifts in the initial number of neutrons stored in the chamber.

Just after filling the chamber with UCN and closing the shutter, during the storage phase, the neutrons still emerging from the source are directly guided to the UCN detectors for a constant time period. These neutrons are referred to as monitor counts. We use monitor counts to normalize the emptying counts as shown in Fig.~\ref{fig5-17} (bottom). Eqs. \ref{eq5-32}, \ref{eq5-13}, and \ref{eq5-34} use the number of neutrons counted after storage under various magnetic field configurations. We shall henceforth replace bare counts such as $N_0(t_s)$, $N_{\bf B}(t_s)$, and $N_{-{\bf B}}(t_s)$ with normalized counts $n_0(t_s)$, $n_{\bf B}(t_s)$, and $n_{-{\bf B}}(t_s)$ as follows:
\begin{eqnarray}
n_{\{0,{\bf B},-{\bf B}\}}(t_s) &=& \frac{N^{\text{emptying}}_{\{0,{\bf B},-{\bf B}\}}(t_s)}{N^{\text{monitor}}_{\{0,{\bf B},-{\bf B}\}}(t_s)} \label{eq5-42}.
\end{eqnarray}
Using Eq.~\ref{eq5-42}  transforms Eqs. \ref{eq5-32} and \ref{eq5-34} as follows:
\begin{eqnarray}
E_B &=& \frac{n_0\left(t_s\right)}{n_{\bf B}\left(t_s\right)} - 1 \label{eq5-45}, \\
A_B &=& \frac{\left(n_{\bf B}\left(t_s\right)-n_{\bf{-B}}\left(t_s\right)\right)}{\left(n_{\bf B}\left(t_s\right)+n_{\bf{-B}}\left(t_s\right)\right)} \label{eq5-46}. 
\end{eqnarray}

The monitor counts are typically of the order of a million, far exceeding the emptying counts which are of the order of few tens of thousands. Thus, the uncertainty on the ratio of emptying and monitor counts is mostly dependent on the uncertainty coming from the emptying counts. Studying the uncertainty on the ratio will also help us understand if the monitor counts are an accurate normalizing number.

\begin{figure*}[h]
\centering
\includegraphics[scale=.19]{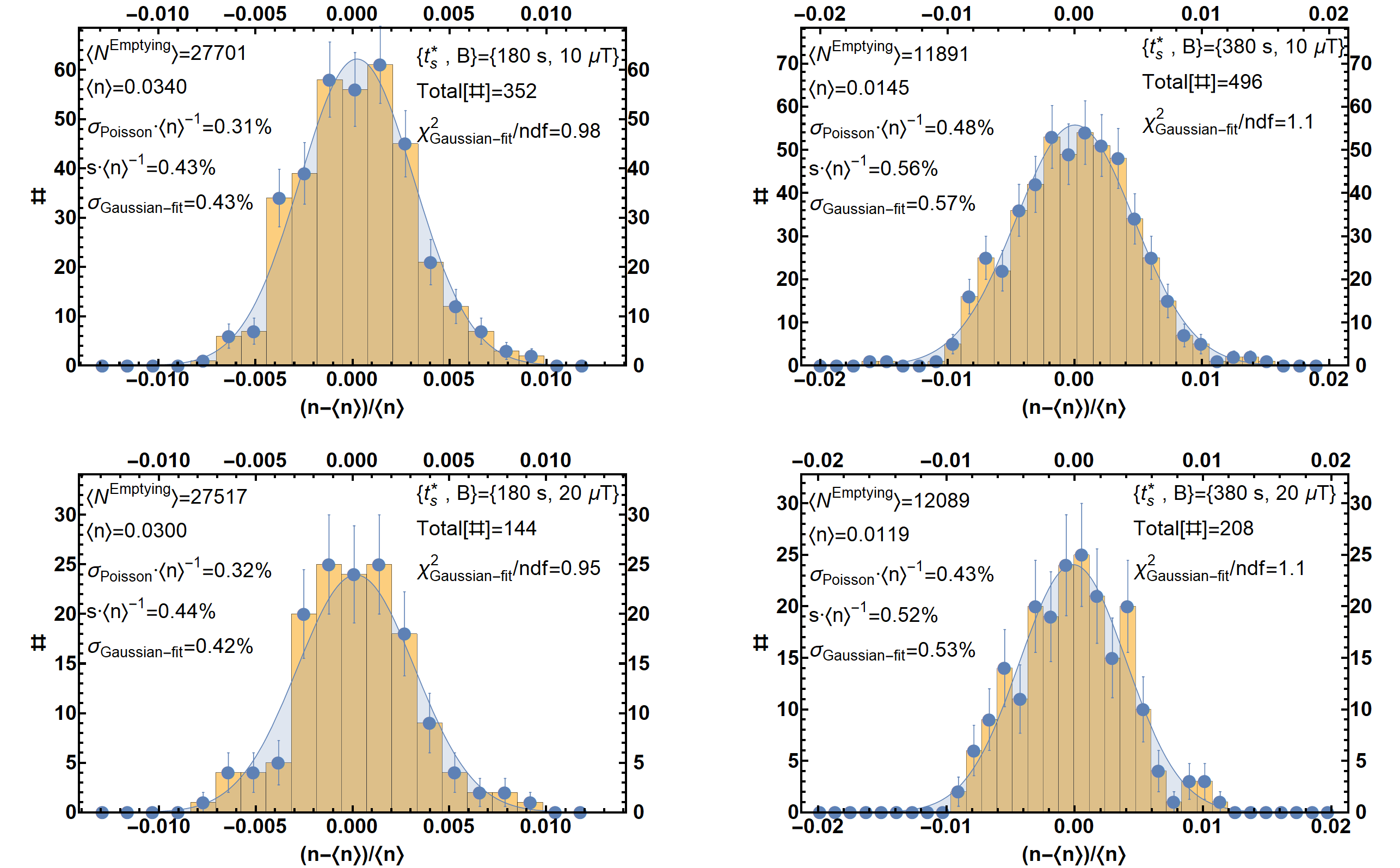}
\caption[]{Histograms of normalized residuals of the ratio between emptying and monitor counts from Fig.~\ref{fig5-17} (bottom), along with the corresponding best fit with a Gaussian distribution. In clockwise order:  top-left corresponds to dwells with run identifiers of $t^*_s = 180~\text{s and}~B_0~\sim10~\mu\text{T}$; top-right corresponds to dwells with run identifiers of $t^*_s = 380~\text{s and}~B_0~\sim10~\mu\text{T}$; bottom-right corresponds to dwells with run identifiers of $t^*_s = 380~\text{s and}~B_0 \sim20~\mu\text{T}$; bottom-left corresponds to dwells with run identifiers of $t^*_s = 180~\text{s and}~B_0 \sim20~\mu\text{T}$. The extracted values are explained in the text.}
\label{fig5-18}
\end{figure*}

Figure~\ref{fig5-17} (top) shows monitor and emptying counts averaged over a single dwell, of 4 cycles. Figure~\ref{fig5-17} (bottom) shows the corresponding ratio between emptying and monitor counts. In the following paragraphs we analyze the relative spread of the normalized counts and the contribution of the Poisson statistical uncertainty of the associated emptying counts. In Fig.~\ref{fig5-17} (bottom), for each of the magnetic field series (in pink regions labeled `$B\sim10~\mu$T',  and green regions labeled `$B\sim20~\mu$T', respectively), we see two distinct levels. The higher level corresponds to $180$ s of micro-timer storage, and the lower level corresponds to $380$ s of micro-timer storage. 
The distribution of the residuals, after subtracting the mean value from the ratio between the UCN counts and their corresponding monitor counts, $(n-\left<n\right>)/\left<n\right>$, is shown in Fig.~\ref{fig5-18} for each of the four series along with a best fit with a Gaussian distribution. Building histograms from normalized residuals allows us to study the relative width parameters of the corresponding distributions (\emph{w.r.t}  $\left<n\right>$). The final analysis will focus instead on the distributions of $E_B$ and $A_B$.

The aim of the present study was to check if the errors propagated from the neutron counts are compatible with the data scatter, and whether the histograms are normal distributions. For each histogram in Fig.~\ref{fig5-18}, we give the average counts per cycle $\left<N^{\text{emptying}}\right>$ and the simple mean of the normalized counts (indiscernible from the weighted mean because of the large counts).  The following width parameters were calculated for comparison: (i) standard deviation, $s$ of $\left<n\right>$ quantifying the scatter of the central values, (ii) $\sigma_{\text{Poisson}}$, the width parameter of the distribution calculated from counting statistics, (iii) width parameter of the best fit with a normal distribution, $\sigma_{\text{Gaussian-fit}}$ along with the associated $\chi^2/ndf$. The corresponding values inserted in Fig.~\ref{fig5-18} were normalized with $\left<n\right>$ for easier interpretation.

Thus for a series of data points $\{n_i, \sigma_{n_i}\}$ we have:
\begin{eqnarray}
\left<n\right> &=& \sum^{\#}_{i=1} \frac{n_i}{\#}, \label{eq5-46-0}\\
s &=& \sqrt{\sum^{\#}_{i=1} \frac{1}{(\#-1)}\left(n_i-\left<n\right>\right)^2} \label{eq5-46-2},\\
\sigma_{\text{Poisson}} &=& \sigma_{\left<n\right>}\sqrt{\#}  \nonumber \\ &=& \sqrt{\frac{\#}{\sum^{\#}_{i=1}(1/\sigma^2_{n_i})}} \label{eq5-46-1}.
\end{eqnarray}
In the case of $\sigma_{\text{Poisson}}$ in Eq.~\ref{eq5-46-1}, the errors associated ($\sigma_{n_i}$) with each ratio $n_i=N_i^{\text{emptying}}/N_i^{\text{monitor}}$, came directly from Poisson statistic uncertainties:
\begin{eqnarray}
\left(\frac{\sigma_{n_i}}{n_i}\right)^2 &=& \left(\frac{\sqrt{N^{\text{emptying}}}}{N^{\text{emptying}}}\right)^2 + \left(\frac{\sqrt{N^{\text{monitor}}}}{N^{\text{monitor}}}\right)^2 . \label{eq5-46-3}
\end{eqnarray}
We found that $s/\left<n\right> \cong \sigma_{\text{Gaussian-fit}}$ with a relatively good $\chi^2$, indicating a normal distribution. 
Note that the Poisson-propagated widths in Fig.~\ref{fig5-18} are dominated by errors arising from emptying counts. The Poisson propagated widths, normalized with $\left<n\right>$, are around a factor of two narrower compared to the widths which one would obtain from $1/\sqrt{\left<N^{\text{emptying}}\right>}$ alone.  For example, the Poisson-propagated width in Fig.~\ref{fig5-18} (top-left) is close to $\sqrt{\left<N^{\text{emptying}}\right>}/\left<N^{\text{emptying}}\right> \divisionsymbol 2 \approx 0.6\% \divisionsymbol 2 \approx 0.3 \%$ (\emph{w.r.t} $\left<N^{\text{emptying}}\right>$). Here, a factor of approximately 2 arises because we are considering ratios, `$n$', which are from emptying and monitor counts from the average of 4 cycles in a dwell.

In all the four series of measurements in Fig.~\ref{fig5-18}, the scatter of the central values expressed by $s$ is larger than $\sigma_{\text{Poisson}}$. This means that the Poisson statistics from the counts alone does not explain the width of these distributions. Indeed, we observed a correlation between the position parameter of the UCN switch and $N^{\text{emptying}}$. The corresponding difference, in quadrature, between $\sigma_{\text{Poisson}}$ and $s$, is consistently $\sim0.3\%$ for all the four sets of runs shown in Fig.~\ref{fig5-18}. 
Based on this histogram analysis, we decided not to use Poisson propagated errors later in the evaluation of the $E_B$ and $A_B$ distributions. Instead we will use the scatter of the center values.

\section{Magnetic field calibration using atomic magnetometers}
During the storage phase, the neutrons are exposed to a magnetic field $B_0$. The PSI nEDM experiment ran with a routine magnetic field of $1~\mu$T and associated linear spatial gradient $|g_z| < 40$~pT/cm. A $B_0$ coil current of $\sim17~$mA is required to produce a $1~\mu$T field. From Refs.\cite{[4],[4-2],[6]} it is clear that we need a magnetic field up to $20~\mu$T for a $n-n'$ oscillation search. $B_0 = 20~\mu$T required a coil current of $\sim340~$mA. We limited the maximum applied magnetic field to $B_0 \le 20~\mu$T, in order to be able to effectively degauss the passive $\mu$-metal magnetic shield.

In the $n-n'$ oscillation search, patterns of magnetic field such as [$0\uparrow0\downarrow0\downarrow0\uparrow0\downarrow0\uparrow0\uparrow0\downarrow$] (where the arrow shows the direction of the $B_0$ vector \emph{w.r.t.} to the vertical direction
) were applied. This magnetic field pattern allows to measure both the ratio and asymmetry values of $n-n'$ oscillations while compensating for drifts in magnetic field. 

The requirement for precision on the magnetic field comes from the condition for the zero mirror field case $\omega \left<t_f\right> \ll 1$ (where $\omega = 45.81~(\mu\text{T.s})^{-1}\cdot B$, and $B$ is the applied magnetic field) under which Eq.~\ref{eq5-13} is valid. Assuming a maximum value of $\left<t_f\right> \sim (0.074 \pm 0.004)~$s obtained from the free flight time in the nEDM chamber in simulations with MCUCN \cite{[9-0]}, we require a precision on the magnetic field better than $\sim0.27~\mu$T (at 1 $\sigma$ C.L.) \cite{[9]} or $\sim0.26~\mu$T (at 95\% C.L.). The power source has a precision better than $\sim0.2~\%$ (relative to $17~$mA which is used to create a $1~\mu$T $B_0$ field), and results in a precision of better than $\sim20~n$T.  This is well within the $\sim0.26~\mu$T requirement. In order to measure a magnetic field with a precision better than $\sim0.26~\mu$T, we do not need to run the $^{199}$Hg co-magnetometer during every cycle. Using a nanoampere meter to measure the current supplied to the $B_0$ coil is sufficient to measure the $B_0$ magnetic field. The analogous condition under which Eq.~\ref{eq5-32} is valid, \emph{i.e.} $\omega' \left<t_f\right> \gg 1$, imposes a lower limit on the validity of the constraints on $\tau_{nn'}$ from this measurement. Assuming a minimum value of $\left<t_f\right> \sim (0.061 \pm 0.003)~$s based on simulations, our results are valid in the range $B' > 0.36~\mu$T (at 1 $\sigma$ C.L.) or $B' > 0.38~\mu$T (at 95\% C.L.). The requirement of precision on the magnetic field better than $\sim0.26~\mu$T also sets the requirement on the linear spatial gradient $|g_z| < 21.67~$nT/cm. 

The magnetic field, $B_0$ can be obtained using the value of the applied coil current. In order to calibrate the current source and to characterize the current and the resulting $B_0$ magnetic field, we used the $^{199}$Hg co-magnetometer \cite{[9-1-1],[9-1-2],[9-1-3]} and $^{133}$Cs magnetometers \cite{[11-1],[11-2]}.  
\begin{figure}[h!]
\centering
\includegraphics[scale=.45]{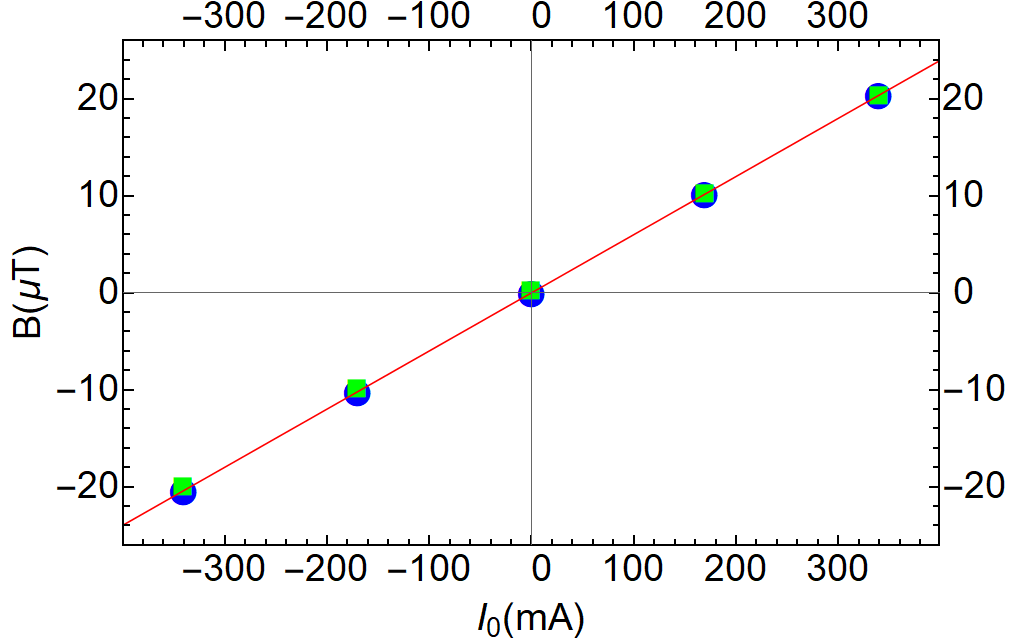}
\caption[]{Plot showing the $B_0$ magnetic field measured by $^{199}$Hg (indicated by blue dots) and $^{133}$Cs (indicated by green squares) magnetometers as a function of $B_0$ coil current.}
\label{fig5-19}
\end{figure}

For the calibration of the dependence of $B_0$ on the coil current, we measured the precession frequencies of $^{199}$Hg and $^{133}$Cs magnetometers. We know the gyromagnetic ratios of atomic $^{199}$Hg and $^{133}$Cs to be $\gamma_{^{199}\text{Hg}}/2\pi = 7.5901152(62)~\text{MHz/T}$ \cite{[10]} and $\gamma_{^{133}\text{Cs}}/2\pi = 3.49862111(39)~\text{GHz/T}$ \cite{[11-2]}. The fit in Fig.~\ref{fig5-19} gives a consistent relationship for the two magnetometer data, between $B_0$ and the coil current:

\begin{eqnarray}
\frac{B_0}{\mu\text{T}}~=~0.05996~(1)~\frac{I_0}{\text{mA}}~+~0.0016~(1) \label{eq5-48}.
\end{eqnarray}

\begin{figure}[t]
\centering
\includegraphics[scale=.45]{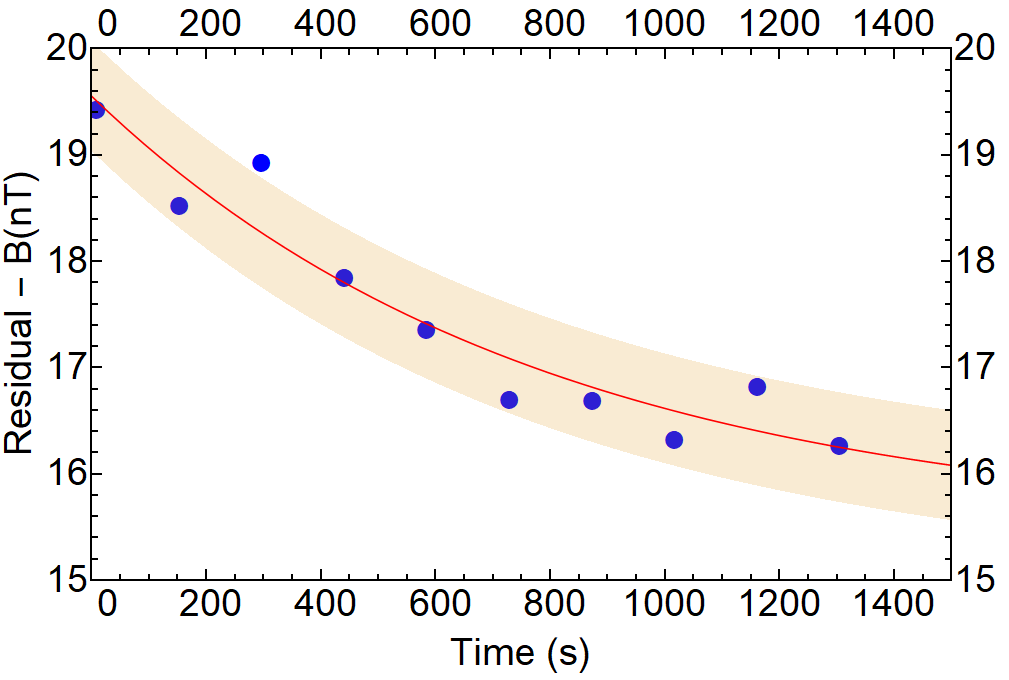}
\caption[]{Decay in the magnitude of $B_0$ magnetic field when the current through the $B_0$ coil is turned off, as measured by the $^{199}$Hg magnetometer. The line and the shaded region indicate an exponential decay fit and its $1 \sigma$ uncertainty, respectively.}
\label{fig5-21}
\end{figure}
Since we did not degauss the $\mu$-metal shield while the magnetic field patterns were applied, we also studied how the $B_0$ magnetic field behaved when it was ramped down in magnitude. When the current to the $B_0$ coil was switched off for the next storage phase of the neutrons in the chamber, it is vital that the magnetic field  is smaller than $0.26~\mu$T so that the condition of $\omega \left<t_f\right> \ll 1$ holds true. In order to measure the residual magnetic field when the current to the $B_0$ coil was switched off, we ramped the magnetic field to $20~\mu$T, the maximum field used in the measurements, held this field for $500~$s, and then switched off the current. This was repeated with reversed polarity. The results are plotted in Fig.~\ref{fig5-21}. The average magnetic field felt by the neutrons during storage when the $B_0$ coil current is off can be obtained by averaging the exponential decay curve in Fig.~\ref{fig5-21} in the appropriate time interval. 
For $300$ s long cycles with micro-timer storage time of $180~$s, storage occurs in the time interval of $[30,210]~$s, whereas for $500$ s long cycles with micro-timer storage time of $380~$s, storage occurs in the time interval of $[30,410]~$s. The average ``zero'' magnetic field experienced by the neutrons during storage is well below the $0.26~\mu$T requirement for both $300$~s and $500$~s long cycles, and given by:
\begin{eqnarray}
\left<B_0(t^*_s)\right>^{B_0 \sim 0}_{t^*_s = [30,210]~\text{s}}=\left(18.98 \pm 0.53\right)~n\text{T}\label{eq5-49}\\
\left<B_0(t^*_s)\right>^{B_0 \sim 0}_{t^*_s = [30,410]~\text{s}}=\left(18.58 \pm 0.73\right)~n\text{T}\label{eq5-50}.
\end{eqnarray}

\section{Statistical sensitivity to $n-n'$ oscillations}
The sensitivity of an experiment searching for $n-n'$ oscillations is represented by the time variable, $\tau_{nn'}$ in Eq.~\ref{eq5-6}. The quantities from which $\tau_{nn'}$, or a constraint on it, can be obtained appear in Eqs.~\ref{eq5-32}, \ref{eq5-13}, and \ref{eq5-34}. In the case where we assume $B'=0$, the oscillation time is purely dependent on the time of storage $t_s$, the mean time-of-flight between two consecutive wall collisions $\left<t_f\right>$, and the neutron count ratio $E_B+1$. When we assume the mirror-magnetic field experienced by neutrons to be non-zero (${\bf B'}\ne0$), the sensitivity, also depends on the applied magnetic field ${\bf B}$. The sensitivity rises dramatically when $\bm{B'}=\bm{B}$, \emph{i.e.} when the neutron and mirror-neutron states are degenerate. 

Note that in the ratio channel under both conditions of ${\bf B'}=0$ and ${\bf B'}\ne0$, as well as in the asymmetry channel under the condition of ${\bf B'}\ne0$, the oscillation time $\tau_{nn'}$ depends on $1/\sqrt{A_B}$ or $1/\sqrt{E_B}$. If no deviations from zero are observed in either $E_B$ or $A_B$, then the constraint on the oscillation time depends on the uncertainty associated with $E_B$ and $A_B$. If one assumes similar individual neutron counts involved in the computation of $E_B$ and $A_B$, \emph{i.e.} $N_0 \sim N_B \sim N_{-B} \approx N$, then the uncertainty on $E_B$ and $A_B$ is given by $2/\sqrt{N}$ and $4/\sqrt{N}$ respectively. Along with Eqs. \ref{eq5-32} and \ref{eq5-13} it follows that:

\begin{eqnarray}
\zeta_{\tau^{{\bf B'=0}}_{nn'}} &\propto& \sqrt[4]{N}\cdot\sqrt{\frac{t_s \left<t^2_f\right>}{\left<t_f\right>}} \qquad \text{\cite{[9]}}  \label{eq5-37}\\ 
\zeta_{\tau^{{\bf B'\ne0}}_{nn'}} &\propto& \sqrt[4]{N}\cdot\sqrt{\frac{t_s}{\left<t_f\right>}}\cdot f(\eta) \label{eq5-38}, 
\end{eqnarray}
where $\zeta$ is the maximum constraint on oscillation time $\tau_{nn'}$, and $f(\eta)$ is a function in units of time. The above equations also imply that an uncertainty on the $\left<t_f\right>$ and the $\left<t^2_f\right>$ distributions, due to uncertainty on the energy spectrum, will affect the final sensitivity by error propagation. This will be treated carefully in the analysis. Here, we consider the $\left<t_f\right>$ variables as fixed and thus constrain ourselves to the statistical sensitivity coming from the spread in UCN counts. 

Null-hypothesis constraints can directly be estimated from Eqs. \ref{eq5-37} and  \ref{eq5-38}, imposing a lower limit on $\tau_{nn'}$ with given statistical confidence, for example $1\sigma \sim 68.3\%$ C.L.. Note that in order to increase the sensitivity, \emph{i.e.} increase the lower bound on $\tau_{nn'}$, we could:
\begin{enumerate}
\item{increase the storage time of the UCNs}
\item{increase the number of neutrons surviving after a storage time of $t_s$}
\item{in the case of $\bm{B'}=0$, use lower-energy UCNs so that the free flight-time is as high as possible (from Eq. \ref{eq5-37}); in the case of $\bm{B'}\ne0$ use high energy UCNs so that $\left<t_f\right>$ is as low as possible (from Eq. \ref{eq5-38}).}
\end{enumerate}
Usually, the sensitivity is a compromise between the above three items, as:
\begin{enumerate}
\item{since the sensitivity always goes as $\sqrt[4]{N}\cdot\sqrt{t_s}$ and neutron counts decay exponentially with increase in storage time, there will be an optimal $t_s$,}
\item{the mean energy of the UCNs decreases with an increase in storage time, resulting in an increase of $\left<t_f\right>$.}
\end{enumerate}

\begin{figure}[h!]
\centering
\includegraphics[scale=0.65]{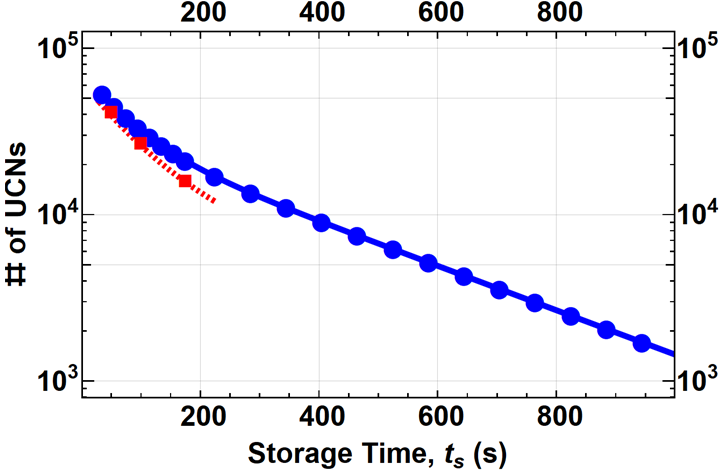}
\caption[]{Number of neutrons as a function of storage time in the PSI nEDM apparatus compared to the storage curve of Ref. \cite{[2]} at the ILL. Solid blue curve, with blue dots for data points, indicates the storage curve measured at PSI for unpolarized UCNs. The dashed red curve, with red squares for data points, shows the storage curve measured at ILL for unpolarized UCNs.}
\label{fig5-9}
\end{figure}
The energy spectra of the UCNs at a specific storage time is a given. However we can optimize the time of storage $t_s$ in view of achieving the best sensitivity, by studying the decay curve of the neutrons in the chamber. In Fig.~\ref{fig5-9}, the decay curve is shown for unpolarized neutrons in the PSI nEDM apparatus,  compared to the measurement at ILL \cite{[2]}. At ILL the measurements routinely counted around $\sim17,000$ neutrons after $180$ s of storage, whilst at PSI we measured on average over $\sim27,000$ neutrons after $180$ s of storage. The number of cycles scales linearly with the number of neutrons counted, thus an increase in the number of cycles has a similar effect as that shown by neutron counts in Eq.~\ref{eq5-38} ($\propto \sqrt[4]{N}$). 

\begin{figure}[h!]
\centering
\includegraphics[scale=.55]{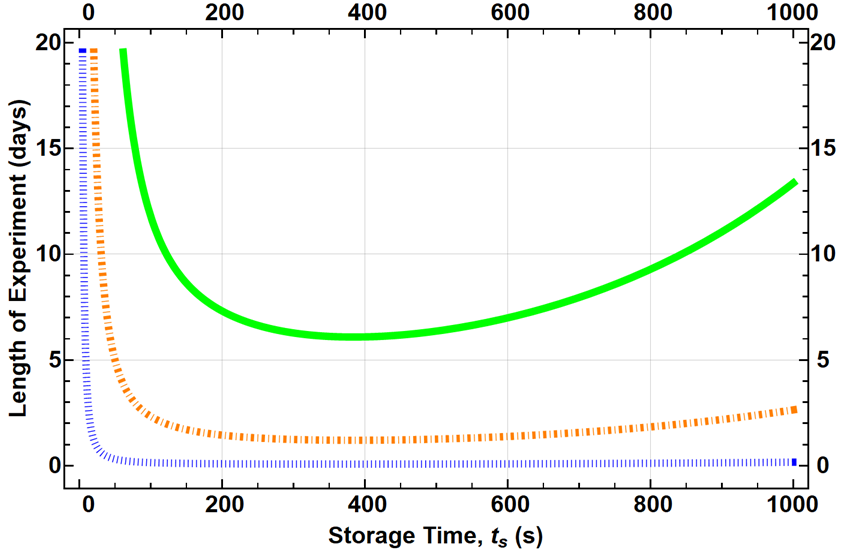}
\caption[]{Plot showing the length of the predicted run time it takes to achieve a sensitivity in $\tau^{\bm{B'}\ne0}_{nn'}$ (at 95\% C.L.) \emph{w.r.t.} storage time. Here, the blue dotted curve indicates a goal of $\tau_{nn'} \ge 12~$s, the orange dot-dashed curve corresponds to a goal of $\tau_{nn'} \ge 24~$s, and the green curve shows the time required to achieve a goal of $\tau_{nn'} \ge 36~$s.}
\label{fig5-11}
\end{figure}
Since this effort is aimed at testing the signals reported in the asymmetry channel (Eq.~\ref{eq5-46}) in Refs. \cite{[2],[3],[5]}, we can estimate the sensitivity of the PSI nEDM apparatus to $\tau_{nn'}$ by scaling up the constraint of $\tau_{nn'} > 12~\text{s}~\forall~(0 \le B' \le 12.5 \mu$T~at 95\% C.L.) in Ref. \cite{[5]}, by using Eqs.~\ref{eq5-37} and \ref{eq5-38}. The PSI nEDM measurement used a micro-timer storage time of $180$ s with each cycle being $300$ s long. Thus an overhead of $120$ s was reasonable to accommodate filling UCNs into the storage chamber and counting the UCNs after storage for time $t_s$. An overhead of $120$ s along with the storage time $t_s$ allows us to calculate the number of cycles per day which then scales linearly with the number of neutrons counted. Along with the decay curve in Fig.~\ref{fig5-9} and Eq. \ref{eq5-38}, by scaling up the published constraint of Ref. \cite{[5]}, we can plot (Fig.~\ref{fig5-11}) the time it takes to achieve a sensitivity in $\tau^{\bm{B'}\ne0}_{nn'}$ (at 95\% C.L.), \emph{viz.} $\tau^{\bm{B'}\ne0}_{nn'}\ge\{12,24,36\}$ s as a function of storage time. From Fig.~\ref{fig5-11} we conclude that the shortest time it takes to achieve a certain sensitivity is always $t^*_s = 380$ s. Therefore we chose to operate our experiment with $380$ s of micro-timer storage time.

Previous experiments were performed with $\sim180~$s of storage time \cite{[5],[2]}. We performed a part of this effort using this storage time, to ensure that we can  compare old simulations of $\left<t_f\right>$  with recent ones. These values will be used in the analysis even though the storage chamber has since been renewed, and the energy spectrum from the PSI UCN source is not identical to the one at ILL.

\begin{figure}[h!]
\centering
\includegraphics[scale=.52]{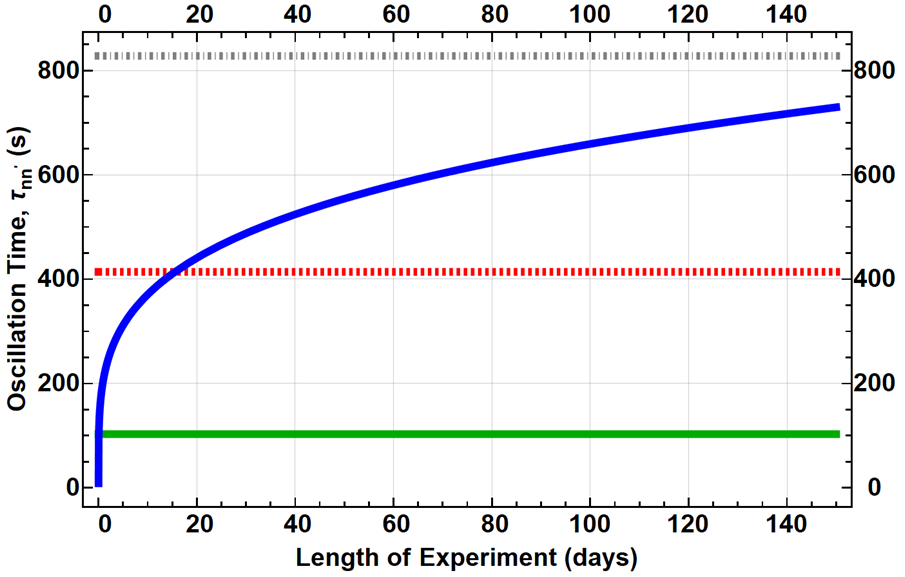}
\caption[]{Plot showing the sensitivity to $\tau^{\bm{B'}=0}_{nn'}$ ($90\%$ C.L.) achievable as a function of total running time of the experiment. The solid blue line marks the sensitivity projection for the PSI nEDM apparatus using 500 s long cycles (with $t^*_s = 380~$s). The dashed lines indicate the constraints imposed by older experiments: the flat dashed red curve shows the constraint from Ref. \cite{[3]}, the flat dot-dashed gray line indicates a constraint which is twice that of the constraint in Ref. \cite{[3]}, and the flat solid green line indicates the constraint from Ref. \cite{[2]}. }
\label{fig5-12}
\end{figure}
In Fig.~\ref{fig5-12} we plotted the sensitivity of the PSI nEDM apparatus to $\tau^{\bm{B'}=0}_{nn'}$ by scaling up the constraint of $\tau_{nn'} > 103$ s (at 95\% C.L.) in Ref. \cite{[2]}. For this we used Eq.~\ref{eq5-37}, and assumed a micro-timer storage time of $380$ s, with a total overhead of $120$ s, resulting in $500$ s cycles. A running time of the order of a month with the PSI nEDM apparatus is sufficient to achieve a sensitivity comparable to the current leading constraint under the assumption of $\bm{B'}=0$ \cite{[3]}. However, Fig.~\ref{fig5-12} also makes it clear that it is unreasonable to schedule for an effective sensitivity a factor two better than the current leading constraints in the PSI nEDM apparatus.

The sensitivity evaluations in Fig.~\ref{fig5-11} and \ref{fig5-12} do not include the contributions from the uncertainty on the energy spectra, affecting $\left<t_f\right>$.

\section{Conclusion}
The measurements by the PSI nEDM collaboration searching for $n-n'$ oscillations were performed using storage times $t^*_s = \{180,380\}~$s corresponding to  total cycle times of $t_t=\{300,500\}~$s, and applied magnetic fields $B_0 = \{10,20\}~\mu$T. These together provided four different configurations for the experiment cycles. While the storage time, $t^*_s = 380~$s is statistically optimized for greatest sensitivity, $t^*_s = 180~$s provides ample performance overlap with the previous generation of experiments reported in Refs.~\cite{[2],[5]}. The $B_0 = \{10,20\}~\mu$T configurations provide ample coverage of the relevant parameter space to test the signals reported in Refs.~\cite{[4],[4-2],[6]}. UCN storage curve data were collected by applying magnetic field patterns of [$0\uparrow0\downarrow0\downarrow0\uparrow0\downarrow0\uparrow0\uparrow0\downarrow$] for $\gtrsim2000$ dwells over 42 days. The long run schedule also allows us to study periodic behavior, which may be interpreted in the $n-n'$ oscillation framework. This provided a sensitivity to $n-n'$ oscillations comparable to the leading constraints for $B'=0$. The final analysis will pay special attention to the uncertainty in the energy spectra of UCNs. This PSI nEDM effort also marks the first dedicated simultaneous search for $n-n'$ oscillations in both the asymmetry and ratio channels.

\section*{Acknowledgments}
The authors thank the exceptional support provided by Michael Meier, Fritz Burri and the BSQ group at PSI. LPC and LPSC groups are supported by ANR grant \# ANR-09-BLAN-0046. University of Sussex group is supported by STFC grants \# ST/N504452/1, ST/M003426/1, and ST/N000307/1. University of Sussex group is also supported by their School of Mathematical and Physical Sciences. PSI group is supported by SNSF grants \# 200020-137664, \# 200021-117696, \# 200020-144473, \# 200021-126562, \# 200020-163413 and \# 200021-157079. ETHZ is supported SNSF grant \# 200020-172639. University of Fribourg group is supported by SNSF grant \# 200020-140421. The Polish groups are supported by National Science Center grant \# 2015/18/M/ST2/00056. For the KU Leuven group, this work is also partly supported by Project GOA/2010/10 and  Fund for Scientific Research in Flanders (FWO). One of the authors, P.~M. would like to acknowledge support from the SERI-FCS award \# 2015.0594 and Sigma Xi grant \# G2017100190747806. We also acknowledge the grid computing resource provided by PL-GRID \cite{[12]}. We would also like to thank Zurab Berezhiani for discussions and theoretical guidance.

\end{document}